\documentclass[%
superscriptaddress,
frontmatterverbose, 
bibnotes,
 amsmath,amssymb,
 aps,
floatfix,
]{revtex4-2}

\usepackage{graphicx}
\usepackage{epstopdf, epsfig}
\usepackage[percent]{overpic}

\usepackage{blindtext}
\usepackage{graphicx}
\usepackage{float}
\usepackage{bm}
\usepackage{multirow}
\usepackage{xcolor}

\usepackage[utf8]{inputenc}
\usepackage{xcolor}
\usepackage{float}
\usepackage{caption}
\usepackage{subcaption}

\begin{document}

\preprint{APS/123-QED}

\title{Rheology of dense  fiber suspensions: Origin of yield stress, shear thinning and normal stress differences}

\author{Monsurul Khan}
 \affiliation{%
 Department of Mechanical Engineering, Purdue University, IN 47905, USA
}
\author{Rishabh V. More}%
\affiliation{%
 Department of Mechanical Engineering, Purdue University, IN 47905, USA
}%
\author{Luca Brandt}

\affiliation{
Linn´e Flow Centre and SeRC (Swedish e-Science Research Centre), KTH Mechanics, SE 100 44 Stockholm, Sweden
}
\author{Arezoo M. Ardekani}
\affiliation{Department of Mechanical Engineering, Purdue University, IN 47905, USA
}%

\date{\today}

\begin{abstract}

We explain the origins of yield stress, shear-thinning, and normal stress differences in rigid fiber suspensions. We investigate the interplay between the hydrodynamic, colloidal attractive and repulsive, and inter-fiber contact interactions. The shear-thinning viscosity and  finite yield stress obtained from the computational model are in quantitative agreement with experiential results from the literature. In this study, we show that attractive interactions result in yield stress and shear-thinning  rheology in the suspensions of rigid fibers. 
 This is an important finding, given the ongoing discussion regarding the origin of the yield stress for suspensions of fibers. The ability of the proposed model to quantitatively predict the rheology is not limited to only shear thinning and yield stress but also extends to normal stresses. 
 

\end{abstract}

\maketitle

\section{Introduction}
Fiber suspensions are widely encountered in natural and industrial applications, with examples in paper and pulp production, biomass solutions, and chemical processing \citep{bivins2005new, elgaddafi2012settling,hassanpour2012lightweight,lundell2011fluid,lindstrom2008simulation}. Under shear, these suspensions display several non-Newtonian properties such as the Weissenberg effect \citep{nawab1958viscosity, mewis1974rheological}, shear thinning \citep{kitano1981rheology, goto1986flow,  bounoua2016shear}, non-zero normal stresses \citep{ snook2014normal, keshtkar2009rheological}, yield stress \citep{bounoua2016shear}. Specifically, shear thinning in fiber suspensions has been an active area of research, and consequently, the literature provides many phenomenological explanations, such as the increase in the effective particle size due to the presence of electric double layer \citep{quemada2002energy}, excluded volume interactions between rigid fibers \citep{raghavan2012conundrum}, elastic bending of flexible micron-sized or nano-sized fibers \citep{bennington1990yield,song2005influence}, and fiber aggregation \cite{ma2008rheological} and nonlinear lubrication force \citep{natale2014rheological}. However, numerical models incorporating these phenomenological explanations are missing from the literature.

Non-Newtonian rheology arises from a variety of interactions such as hydrodynamics, cohesive \citep{bounoua2016normal,singh2019yielding}, contact \citep{khan2021rheology,lobry2019shear}, and is influenced by the physical properties of fibers, e.g., roughness, shape, size distribution, etc. \citep{wu2010numerical, banaei2020numerical} Each of these interactions leads to a corresponding stress scale as ${F}_A/\pi d^2$ (d is the fiber characteristic scale) that is competing with hydrodynamic interactions that scale as $\pi \eta \dot{\gamma} d^2$, where $\dot{\gamma}$ is the imposed shear rate and $\eta$ is the viscosity of the suspending fluid. The competition between these stress scales could lead to a rate-dependent rheological behavior \citep{more2020unifying,guazzelli2018rheology}. Among these different interactions between fibers, the influence of short-range hydrodynamic forces and  direct mechanical contact has been understood through theoretical modeling \citep{djalili2006fibre,ferec2009modeling} and numerical simulations \citep{khan2021rheology,switzer2003rheology}. However, there is little documentation on the study of fiber suspensions with colloidal interactions, even though van der Waals forces \citep{maranzano2001effects}, depletion forces due to dissolved non-interacting polymers, \citep{gopalakrishnan2004effect}, presence of external fields \citep{brown2010generality} can lead to attractive forces between fibers. Therefore, it is crucial to understand the effect of attractive and repulsive forces in fiber suspensions as they produce nonlinear scaling of the shear stress with shear rate. Earlier efforts, including experiments on an attractive system with  nanofibers, show that the interplay between electrostatic repulsion and the van der Waals attraction governs the degree of fiber flocculation and yielding behavior \citep{solomon1998rheology, michot2009sol}. Moreover, shear thinning was observed due to adhesive interactions in the suspension of rigid micro-sized rod-like particles such as polyamide (PA) \citep{chaouche2001rheology} and ceramic fibers \citep{bergstrom1998shear}. However, the role of colloidal interactions remained unclear. Lately, a theoretical model of aggregated fiber suspensions, considering the adhesive force between fibers, shows a good match with experiments on the rigid fiber suspensions \citep{bounoua2016shear,bounoua2016apparent}. This model included fitting parameters, which were obtained from experimental data. None of these explanations describe the behavior of normal stresses, and so their applicability is limited.

The apparent yield stress,  the minimum stress required to begin the flow, is considered one of the most important rheological properties of fiber suspensions. The yield stresses in the suspensions rise as the volume fraction increases, and they are more noticeable for greater aspect ratios \citep{tapia2017rheology, keshtkar2009rheological}. A recent theoretical model that considered attractive interactions between fibers in the dilute regime predicted the Bingham law for the shear stress, with apparent yield stress proportional to the square of volume fraction $(\phi^2)$ \citep{bounoua2016apparent}.  The shear stress for suspensions of larger-sized rigid fibers also follows the Bingham law, but the yield stress rises with higher power laws in $\phi$ than predicted in previous studies \citep{tapia2017rheology}. The origin of yield stress has been attributed to adhesive contacts even though the fiber size was large \citep{ chaouche2001rheology}. As pointed out in \cite{tapia2017rheology}, 
recent experimental studies  failed to come to a conclusion on whether the fiber size or the attractive forces are responsible for the yield stress and mentioned the necessity of further investigation to pinpoint the origin of the yield stress.


As there is no clear explanation of the shear thinning and yield stress in fiber suspensions, we provide a predictive model in this paper. Our proposed model incorporates short-range interactions via attractive and repulsive interactions and contact interactions that quantitatively capture shear thinning rheology and yield stress in the fiber suspensions, and elucidate the effect of colloidal interactions. Moreover, our model not only explains the yield stress and shear thinning behavior but also accurately predicts the normal stresses, which further strengthens the validity of our model.
Lastly, we demonstrate the versatility of the proposed model by capturing the effect of changing surface properties by accurately modeling the contact dynamics. 
To the best of the authors' knowledge, this is the first computational study to quantitatively capture the experimentally observable rheological behavior (relative viscosity $\eta_r$, first normal stress coefficient $\alpha_1$, second normal stress coefficient $\alpha_2$, and yield stress $\sigma_y$) for fiber suspensions and demonstrate the underlying physical mechanism. 

\section{Simulation methodology}

\begin{figure}
  \centerline{\includegraphics[width=.5\linewidth]{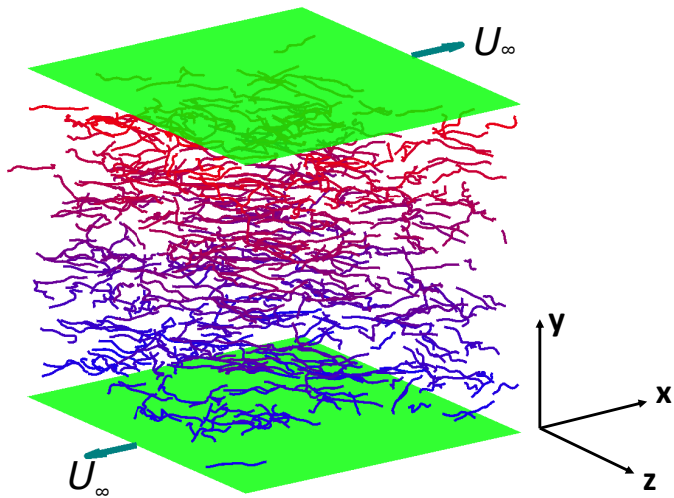}}
  \caption{Simulation setup of the shear flow of a fiber suspension. The top and bottom walls move with velocities $U_{\infty} = \dot{\gamma}L$ in the directions shown by the arrows. $\dot{\gamma}$ is the imposed shear rate, and $L$ is the distance between the walls. }
\label{fig:geometry}
\end{figure}

We perform direct numerical simulations of neutrally buoyant fibers with aspect ratio $AR = l/d$, where $l$ is the length and $d$ is the diameter of the fiber, in a shear flow generated by the top and bottom walls moving in opposite directions with a velocity $U_{\infty}=\dot{\gamma}L$ generating an imposed shear rate  $\dot{\gamma}$, as shown in figure~\ref{fig:geometry}. Here, $L$ is the distance between the walls. 

The suspending fluid is incompressible Newtonian with a viscosity $\eta$, and its flow is governed by the Navier-Stokes equations.
\begin{equation}
\frac{{\partial \mathbf{u}}}{{\partial t}} + \mathbf{\nabla} \cdot (\mathbf{u} \otimes \mathbf{u}) =  - \nabla p + \frac{1}{{{\mathop{\rm \textit{Re}}\nolimits} }}{\mathbf{\nabla} ^2}\mathbf{u} + \mathbf{f},
\label{NS}
\end{equation}
\begin{equation}
\mathbf{\nabla} \cdot \mathbf{u} = 0,
\end{equation}
where  $\mathbf u$ is the velocity field, $p$ is the pressure, $\mathbf f$ is the volume force to account for the suspending fibers, and $Re = \rho\dot{\gamma}l^2/\eta$ is the Reynolds number, where $\rho$ is the fluid density, $l$ is the characteristic length scale which is also the fiber length. 
We model the fibers as in-extensible slender bodies. So, their motion for the neutrally buoyant case is described by the Euler-Bernoulli beam equation as \citep{banaei2020numerical}: 

\begin{equation}
\frac{{{\partial ^2}\mathbf X}}{{\partial {t^2}}} = \frac{{{\partial ^2}{\mathbf X_{fluid}}}}{{\partial {t^2}}} + \frac{\partial }{{\partial s}}(T\frac{{\partial \mathbf X}}{{\partial s}}) - B\frac{{{\partial ^4}\mathbf X}}{{\partial {s^4}}} - \mathbf F + \mathbf {F}^f,
\label{pinal}
\end{equation}
where $s$ is the curvilinear coordinate along the fiber, $\mathbf X = (x(s,t),y(s,t),z(s,t))$ is the position of the Lagrangian points on the fiber axis, $T$ is the tension, $B$ is the bending rigidity,  $\mathbf F$ is the fluid-solid interaction force, $\mathbf {F}^f$ is the net inter-fiber interaction. The fibers are considered  inextensible, expressed as \citep{huang2007simulation,pinelli2017pelskin} : 
\begin{equation}
 \frac{{\partial \mathbf{X}}}{{\partial s}}.\frac{{\partial \mathbf X}}{{\partial s}} = 1.
 \label{inextensible}
\end{equation}
We use the immersed boundary method (IBM) \citep{peskin1972flow} to couple the motion of fluid and solid fibers. For the details of the numerical method, readers are referred to the supplemental material.

\begin{figure}
 \centerline{\includegraphics[width=0.5\linewidth]{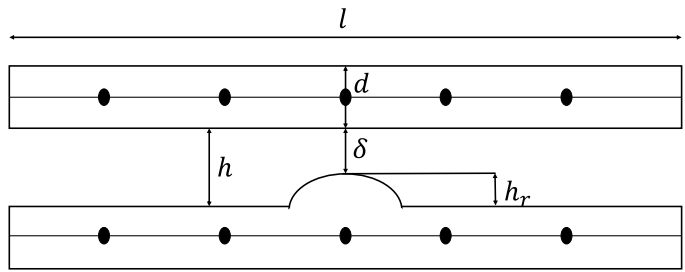}}
  \caption{A sketch of the roughness model, $l$ and $d$ are the length and diameter of the fiber, respectively, $h_r$ is the asperity height, and $\delta=h-h_r$ is the surface overlap. Contact occurs when $\delta \le 0$. Dots along the axes of the fibers indicate Lagrangian points.}
  \label{fig:deformation}
\end{figure}

In the numerical simulation, the hydrodynamic interactions are well resolved with the IBM. However, a fine Eulerian mesh is required to capture the short-range interactions that increase the computational cost. So, we use the proposed model to calculate the short-range interactions. The short-range interaction, $\mathbf F^f=\mathbf F^{lc}+ \mathbf F^c+ \mathbf{F}^{cons} $, is split into the lubrication correction $\mathbf{F}^{lc}$, contact force $\mathbf{F}^c$, conservative force $\mathbf{F}^{cons}=\mathbf{F}^A+\mathbf{F}^R$, where $\mathbf{F}^A$ is the van der Waals attractive force, and $\mathbf{F}^R$ is the repulsive force of electrostatic origin. The implementation of the lubrication correction $\mathbf{F}^{lc}$ can be found in the supplemental
material. The expressions for attractive and repulsive interactions are readily available from theoretical analyses and previous experimental data \citep{bergstrom1998shear, israelachvili2011intermolecular}. The attractive force of van der Waals origin acts in the normal directions toward the fibers and is modeled as $\mathbf F^A = F_A/(h^2+H^2)$, where $h$ is the inter-fiber surface separation. Moreover,  H is fixed to 0.01 to prevent the divergence in $\mathbf F^A$ when $h \rightarrow 0$ (during contact). The strength of the attraction is controlled by $F_A$, which determines the values of the attractive force in contact. The repulsive force $\mathbf{F}^R$ also acts in the normal direction to the fibers but is opposite to the attractive force. $\mathbf{F}^R$ decays with the inter-fiber separation $h$ over the Debye length $\kappa $ as $\mathbf F^R = F_R\exp(-h/\kappa)$ \cite{singh2019yielding,more2020unifying}. The contact between the fibers occurs when the inter-fiber separation distance $h$ becomes smaller than the height of surface asperity $h_R$ as shown in figure \ref{fig:deformation}. Specifically, the single-asperity model of the surface roughness has been widely used owing to its simplicity and effectiveness \citep{gallier2014rheology, lobry2019shear, more2020constitutive, more2020roughness, more2020effect}. Hence, we take the same approach and model the asperity as a hemispherical bump on the fiber surface. Actual asperities might not be just hemispherical and can come in various geometries \citep{tanner2016particle}. However, on average, we can model their behavior by approximately assuming them hemispherical as routinely done in the tribology literature \citep{broedersz2014modeling}. Finally, we split the contact force $\mathbf F^c$ into the tangential ($\mathbf{F}_t^C$) and normal ($\mathbf{F}_n^C$) components. The normal contact force is modelled using a Hertz law, $|\mathbf{F}_n^C|=- k_n|\delta|^{3/2} $, where $\delta = h_R - h$ is the asperity deformation, and $k_n$ is the normal stiffness, which is a function of the fiber material properties \citep{lobry2019shear, more2020effect}.   The Coulomb's friction law gives the tangential force, $|\mathbf F_t|=\mu|\mathbf F_n|$, where $\mu = 0.4$ in the current work unless mentioned otherwise. The typical value of $\mu$ is 
$0.3-0.5$ as measured experimentally for the polymer fibers \citep{bowden1964friction,petrich1998interactions}. Even though little is known about the friction mechanics at the nanoscale, the Coulomb friction allows a correct prediction for flows of colloidal nanoparticles \citep{fujita2008simulation} or hard-sphere suspensions \citep{gallier2014rheology}. We use a repulsive force magnitude as the characteristic force to scale the various forces. Consequently, the characteristics  shear rate scale is $\dot{\gamma}_0={F_R/\pi \eta d^2}$ and the corresponding shear stress scale is $\sigma_0 = F_R/\pi d^2$.

We calculate the bulk stress ($\Sigma_{ij}$) by volume averaging the viscous fluid stress and stress generated by the presence of fibers and inter-fiber interactions. There are three main contributions to the bulk stress: 1) the hydrodynamic contribution, $\Sigma_{ij}^h$ 2) the contact contribution, $\Sigma_{ij}^c$ and 3) the non-contact contribution, $\Sigma_{ij}^{nc}$. The calculation of bulk stress, including different contributions, is described in the supplemental material. 
Rheological properties can be quantified from the bulk stress, e.g., the relative viscosity, $\eta_r = \sigma/\eta\dot{\gamma}$, where $\sigma$ is the total shear stress in the suspension. The first and second normal stress coefficients are defined as $\alpha_1 (\phi) = (\sigma_{xx} - \sigma_{yy})/\sigma$ and $\alpha_2 (\phi) = (\sigma_{yy} - \sigma_{zz})/\sigma$, where $\phi$ is the volume fraction of fibers defined as $\phi  = \frac{{N\pi }}{{4V{(AR)}}^2}$, where $V$ is the volume $(5l\times 8l \times5l)$ of the simulation cell and $N$ is the total number of fibers. We present the time-averaged steady-state values of rheological properties after discarding the initial transient behavior. 
\section{Results and discussion}
\subsection{Shear-rate dependent rheology}
We start our analysis by  demonstrating the accuracy of the proposed model by directly comparing the calculated relative viscosity and shear stress with experimental measurements of neutrally buoyant Polyamide (PA) fiber suspensions, which exhibit a yield stress and shear thinning viscosity \citep{bounoua2016shear,bounoua2016apparent}. PA fibers were suspended in a Newtonian fluid (a mixture of UCON oil 75H90000 in distilled water) at different volume fractions ranging from 1\% to 17\%, depending on the fiber aspect ratio. The roughness ($h_R$) of the fibers has been measured using atomic force microscopy, having values $h_R =5 \pm 2 nm $ and $h_R =14 \pm 4 nm $ for the aspect ratio, $AR$ = 18 and $AR$ = 33, respectively. The dimensionless roughness calculated as $\epsilon_r=h_R/d$ is 0.0003 and 0.001 for aspect ratios 18 and 33, respectively. Moreover, the fibers used in the experiment are almost rigid. The parameters chosen in our simulations mimic these experimental conditions. 

\begin{figure}
  \centerline{\includegraphics[width=1.0\linewidth]{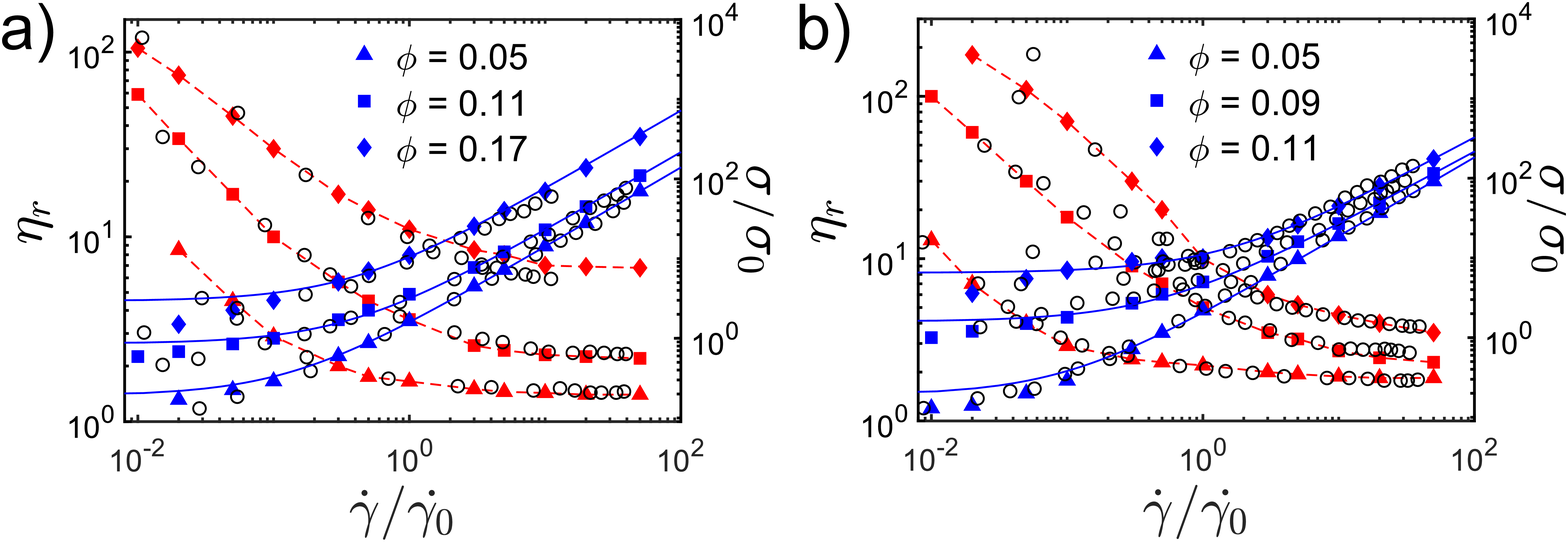}}
  \caption{Experimental and numerical comparison of relative viscosity $\eta_r$ (red symbol) and shear stress (blue symbol) versus non-dimensional shear rate at different volume fractions:  fibers with aspect ratio (a) $AR = 18$; (b) $AR = 33$. Filled symbols with dashed lines show the numerical data. Hollow symbols correspond to experiments, and solid lines in the shear stress curve denote the best fit with the Herschel–Bulkley model (equation~\ref{equation:herschel}) for numerical data. Experimental shear rate and shear stress has been scaled by $\dot{\gamma_0} = 2.5$ and  ${\sigma_0} = 0.7$, respectively. Inclusion of attractive and repulsive interactions along with inter-fiber
contact interactions quantitatively reproduce the experimentally observed shear-thinning viscosity and shear stress \citep{bounoua2016shear}.}
\label{fig:vis_shear_stress_comparison}
\end{figure}

\begin{figure}
  \centerline{\includegraphics[width=1.0\linewidth]{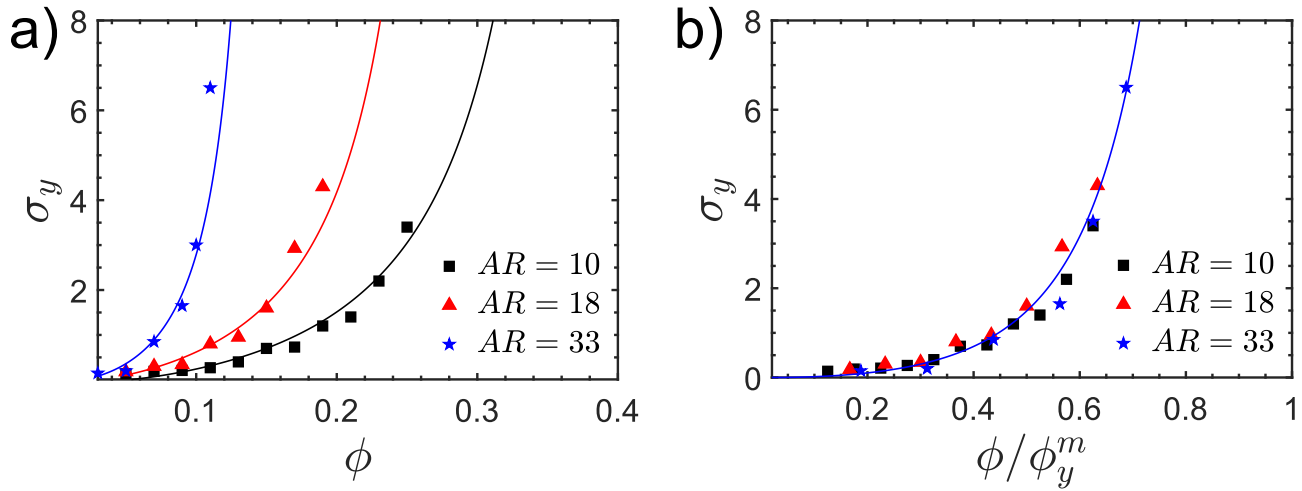}}
  \caption{(a) Yield stress $\sigma_y$ as a function of volume fraction. The solid line shows the best fit of the data to equation \ref{equation:yield}. (b) Yield stress $\sigma_y$ as a function of re-scaled volume fraction. Re-scaling the volume fraction $\phi$ by $\phi_y^m$ leads to the collapse of data indicating that it is the ratio of $\phi$ to $\phi_y^m$  that determines the suspension rheology.}
\label{fig:yield_stress}
\end{figure}

The numerical and experimental comparison of the relative viscosity and flow curves in figures~\ref{fig:vis_shear_stress_comparison}a and ~\ref{fig:vis_shear_stress_comparison}b for the aspect ratios $AR = 18$  and $AR = 33$ shows that the proposed computational model does an excellent job capturing the rate-dependent rheological behavior. We notice that the high shear plateaus of the relative viscosity shift to higher shear rate values as the volume fraction increases. The origin of the shear thinning rheology will be explained in the next section.
From the computational flow curve (blue symbols) in figure~\ref{fig:vis_shear_stress_comparison}, we observe that our numerical model does reproduce the low shear plateau, ($\frac{\dot{\gamma} }{\dot{\gamma}_0} < 0.10$), reminiscent of the yield stress  and the linear segments of the flow curve at high dimensionless shear rates, ($\frac{\dot{\gamma} }{\dot{\gamma}_0} > 0.10$), observed in experiments. Due to this trend in the flow curve, the shear stress of the fiber suspensions can be described by the Herschel–Bulkley model as:
\begin{equation}
    \frac{\sigma}{\sigma_0} = \sigma_y + k(\frac{\dot{\gamma}}{\dot{\gamma}_0})^n
    \label{equation:herschel}
\end{equation}
where $\sigma_y$ is the dimensionless yield stress, $k$ is the consistency index, and $n < 1$ is the shear thinning index. The solid lines in the flow curve (figure~\ref{fig:vis_shear_stress_comparison}) demonstrate the best fit to the Herschel-Bulkley model for the computational data. The yield stress $\sigma_y$ and the fitting parameter are provided in table~\ref{tab:herschel_param}. 
The yield stress for a given volume fraction is  higher for fibers with a higher aspect ratio. Furthermore, as the volume fraction rises, the yield stress increases for a fixed aspect ratio. To analyze the concentration dependence of the yield stress, we plot the yield stress for different volume fractions in figure~\ref{fig:yield_stress}a. It is clear that $\sigma_y$ has a weak dependence on $\phi$ in dilute and semi-dilute regimes ($\phi \le 0.10$ for $AR = 10$ and $\phi \le 0.05$ for $AR$= 18); the dependence is seen to increase more rapidly with $\phi$ for concentrated regimes. The dependence of yield stress $\sigma_y$ on $\phi$ can be well described using the relation proposed by \cite{zhou1995yield}:
\begin{equation}
    {\sigma _y} = {\sigma _c}{\left[ {\frac{{\frac{\phi }{{\phi _y^0}} - 1}}{{1 - \frac{\phi }{{\phi _y^m}}}}} \right]^\frac{1}{\beta} }\
    \label{equation:yield}
\end{equation}
where $\phi_y^0$ is a threshold volume fraction above which the suspension exhibits yield stress. In addition, $\phi_y^m$ denotes the maximum volume fraction at which the yield stress diverges and beyond which the suspension does not move at any applied stress. The fitting parameters are provided in  table~\ref{tab:yield_param}. The threshold volume fraction $\phi_y^0$ at which the suspension first exhibits yield stress and maximum volume fraction $\phi_y^m$ at which the yield stress diverges decreases with the aspect ratio.  Figure~\ref{fig:yield_stress}b shows the data after rescaling the volume fractions of figure~\ref{fig:yield_stress}a by $\phi_y^m$, which leads to the collapse of the data. Hence, the ratio of $\phi$ to the maximum volume fraction $\phi_y^m$ at which the yield stress diverges determines the suspension rheology. Hence, $\phi/\phi_y^m$ can be used as a design parameter for tuning the fiber suspension rheology, as it is easier to measure and control compared to other parameters like size distribution, roughness, and friction in real-world suspensions. 
In the next section, we focus on the origin of shear thinning and yield stress to understand the underlying mechanism and  control the rheological response.

\begin{table}
  \begin{center}
\def~{\hphantom{1.25}}
 \begin{tabular}{|p{4em}|c|c|c|c|}
   \hline
      $ AR $ & $\phi$ & $\sigma_y$   &   $k$  &    $n$  \\[1pt]
       \hline
      \multirow{3}{4em}{$18$} &~ $0.05$~ & ~$0.1898$~ & ~$1.409$~ & ~$0.9975$ ~  \\ 
      \cline{2-5}
      &~ $0.11$~ & ~$0.8571$~ & ~$2.301$~ & ~$0.9865$~ \\
       \cline{2-5}
       &~ $0.17$~ & ~$2.928$~ & ~$7.608$~ & ~$0.9851$~ \\[1pt]
      \hline
       \multirow{3}{4em}{33} &~ $0.05$~ & ~$0.1903$~ & ~$1.882$~ & ~$0.9935$~ \\
       \cline{2-5}
       &~ $0.09$~ & ~$1.587$~ & ~$3.028$~ & ~$0.9255$~ \\
       \cline{2-5}
       &~ $0.11$~ & ~$6.5$~ & ~$4.58$~ & ~$0.9218$~ \\
        \hline
       
  \end{tabular}
  \caption{Herschel–Bulkley parameters (equation \ref{equation:herschel}) for different volume fractions for aspect ratios 18 and 33.}
  \label{tab:herschel_param}
  \end{center}
\end{table}
\begin{table}
  \begin{center}
\def~{\hphantom{1.25}}
  \begin{tabular}{|p{4em}|c|c|c|c|c|}
  \hline
      $AR$  & $\phi_y^0$   &   $\phi_y^m$  &    $\sigma_c$  &$\beta$ \\[1pt]
      \hline
      $10$  &~ $0.05$~ & ~$0.40$~ &~$0.17$~ &~$0.82$~\\
      \hline
      $18$  &~ $0.03$~ & ~$0.30$~ &~$0.1369$~ &~$0.8015$~\\
      \hline
      $33$  &~ $0.017$~ & ~$0.16$~ &~$0.09$~   &~$0.7473$~ \\
      \hline
       
  \end{tabular}
  \caption{Fitting parameters of equation \ref{equation:yield} for aspect ratios, $AR$ = 10, 18 and 33.}
  \label{tab:yield_param}
  \end{center}
\end{table}

\subsection{Origin of shear thinning and yield stress}

\begin{figure}
  \centerline{\includegraphics[width=1.0\linewidth]{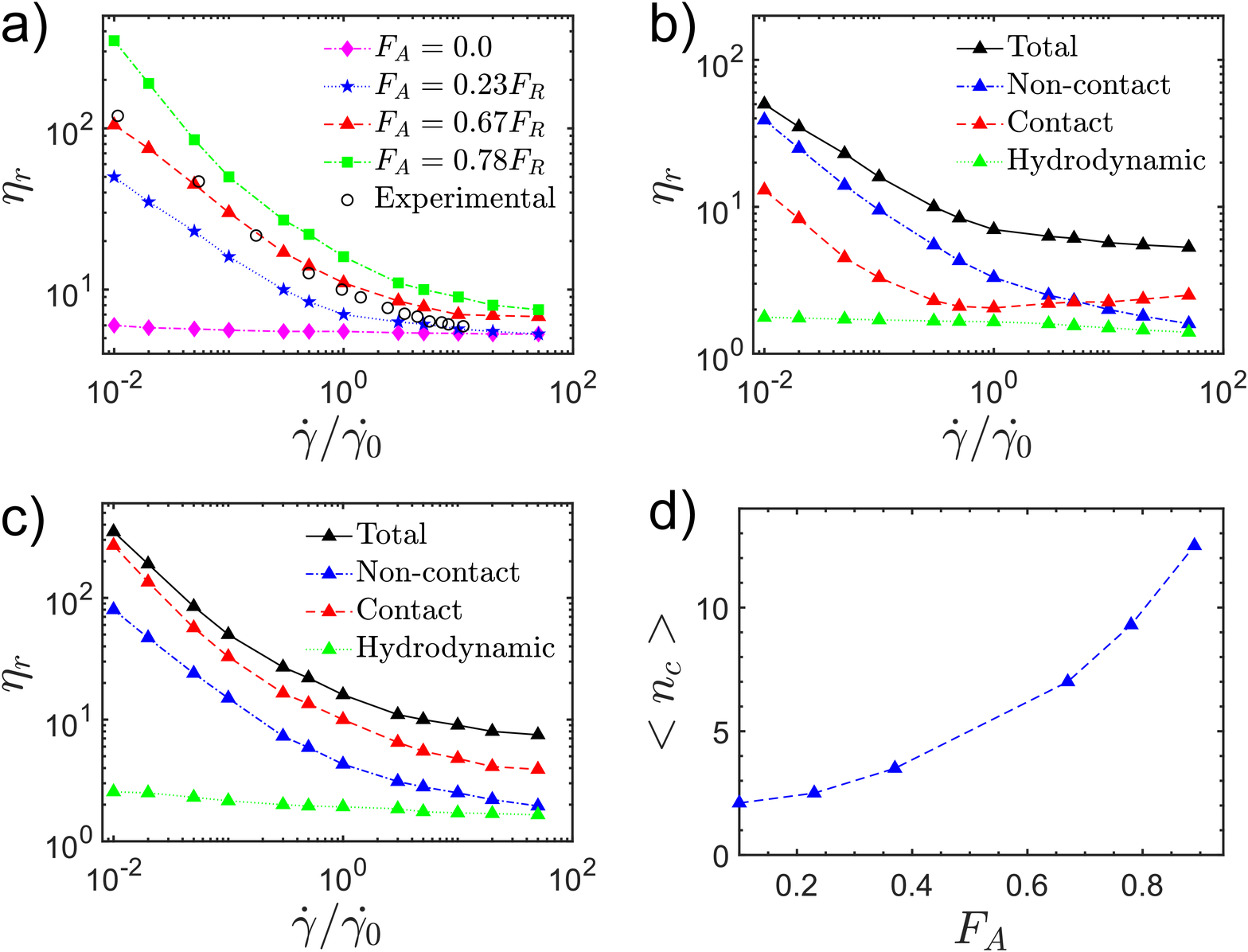}}
  \caption{(a) Relative viscosity plotted as a function of dimensionless shear rate for different values of attractive force $F_A$. Total relative viscosity  and contributions arising from hydrodynamic, conservative, and contact forces, plotted as a function of dimensionless shear rate for $AR = 18$, with (b) $F_A = 0.23F_R$ and (c) $F_A = 0.78F_R$. (d) Average number of fibers in contact as a function of attractive force at $\dot{\gamma}/\dot{\gamma_0} = 0.02$. For all cases, the volume fraction is fixed to $\phi=0.17$. As the attractive force increases, the relative viscosity shifts from being dominated by the non-contact interactions to the frictional contacts induced by the attractive force.}
 \label{fig:viscosity_attractive_conribution}
\end{figure}

We focus on the origin of yield stress, especially with increasing the attractive force, in an attempt to acquire a more mechanistic understanding of the shear-thinning behavior. Figure~\ref{fig:viscosity_attractive_conribution} depicts the effect of varying the attractive force magnitude on the rheology of a fiber suspension. The aspect ratio and volume fraction are fixed to 18 and 17\%, respectively. We do not observe yield stress in the suspension for $F_A = 0$.  
With an increase in the magnitude of the attractive force, the slope of the shear thinning curve at low shear rate increases, which leads to a rise in the yield stress, as shown in figure~\ref{fig:viscosity_attractive_conribution}a. 
With the increase in $F_A$, the separation distance $h$ below which the conservative force is repulsive decreases (supplemental material figure~S3). At a sufficiently high attractive force, the position of zero force and cut-off separation for lubrication ($d/4$) are the same, bringing fibers into direct frictional contact. These frictional contacts can withstand applied shear stress, increasing the yield stress and viscosity. The role of colloidal interactions can be better understood by  separating the hydrodynamic, contact, and non-contact (attractive+repulsive) contributions to viscosity (figures~\ref{fig:viscosity_attractive_conribution}b, \ref{fig:viscosity_attractive_conribution}c). The hydrodynamic contribution to the overall viscosity is negligible for the conditions under consideration. The data plotted here shows an increase in the contact contribution to the relative viscosity as the  attractive force increases. At low attractive force values, the non-contact forces contribute the most to the overall viscosity, while contact contributions take over at higher shear rates. Even though the attractive force is the cause of the yield stress and must be large enough to generate the yield stress, we find that as the attractive force increases, the yield stress shifts from being dominated by the non-contact attraction to frictional contacts generated by the attractive force. Higher attractive forces pull the fibers into more direct contact, as shown in figure~\ref{fig:viscosity_attractive_conribution}d, leading to an increase in the yield stress. 


\begin{figure}
  \centerline{\includegraphics[width=0.7\linewidth]{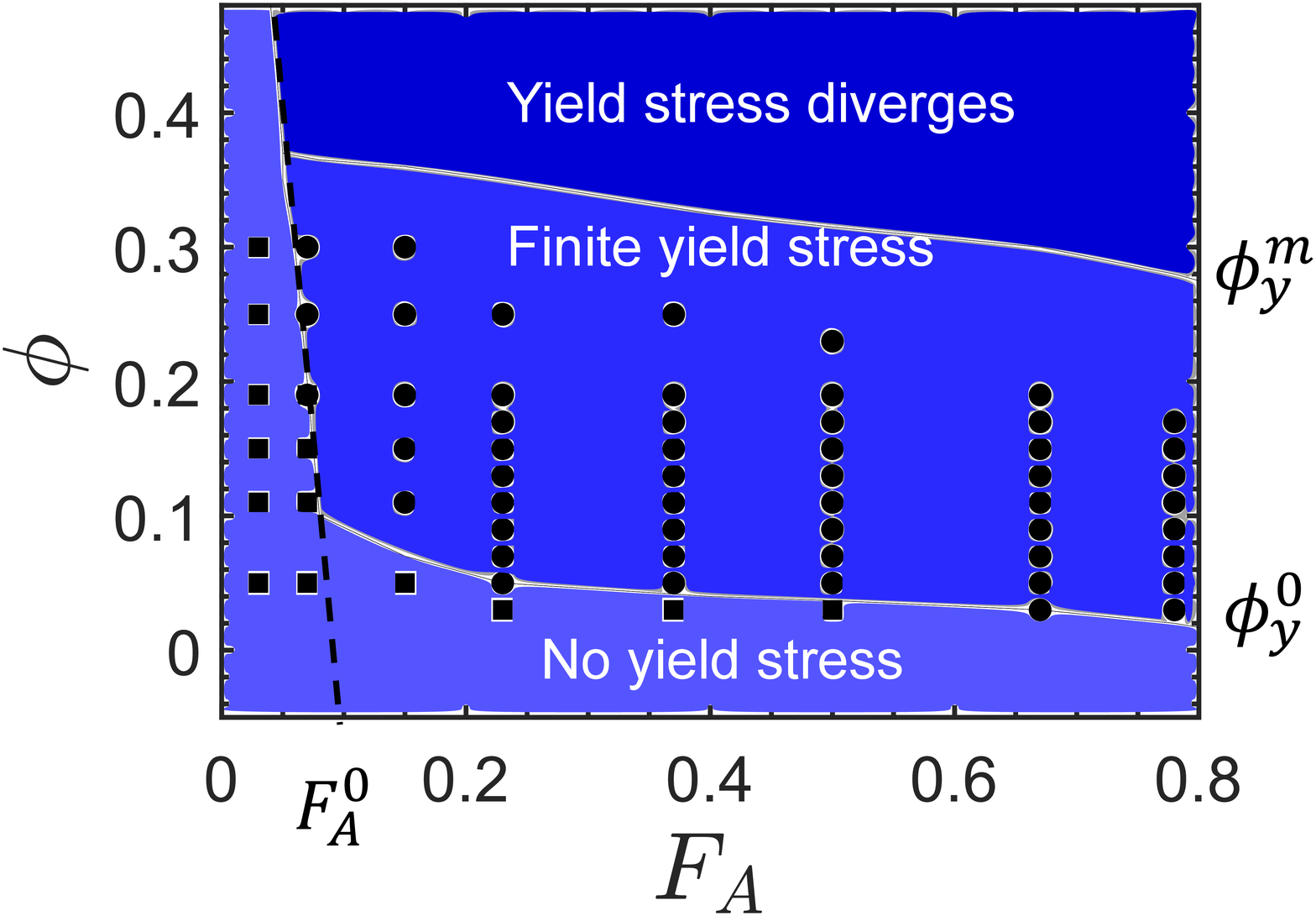}}
  \caption{The flow-state diagram shown for the volume fraction - 
  attractive force plane showing no yield stress (light blue), finite yield stress,  and divergence of the yield stress (dark blue). $\phi_y^0$ and $\phi_y^m$ lines show the limit below which there is no yield stress and above which the yield stress diverges, respectively.} 
\label{fig:flowstate}
\end{figure}

The results of this study show that as we change the volume fraction and the attractive force, the material undergoes different rheological states, which can be demonstrated in a flow-state diagram.
Using the numerical results in the current work, we generate a flow-state diagram in the $\phi-F_A$ plane for  $AR = 18$, as shown in figure~\ref{fig:flowstate}. For the range of attractive forces, the suspension is in different states for $\phi<\phi_y^0$, $\phi_y^0<\phi<\phi_y^m$, and $\phi>\phi_y^m$. Below an attractive force threshold ($F_A^0$), no yield stress is observed at any volume fraction. Above the threshold of attractive force, the system is in the finite yield stress state for $\phi_y^0<\phi<\phi_y^m$. Above $\phi_y^0$, the yield stresses increase with $\phi$ and diverge at $\phi_y^m$.


\subsection{Normal stress coefficients}
\begin{figure}
  \centerline{\includegraphics[width=1.0\linewidth]{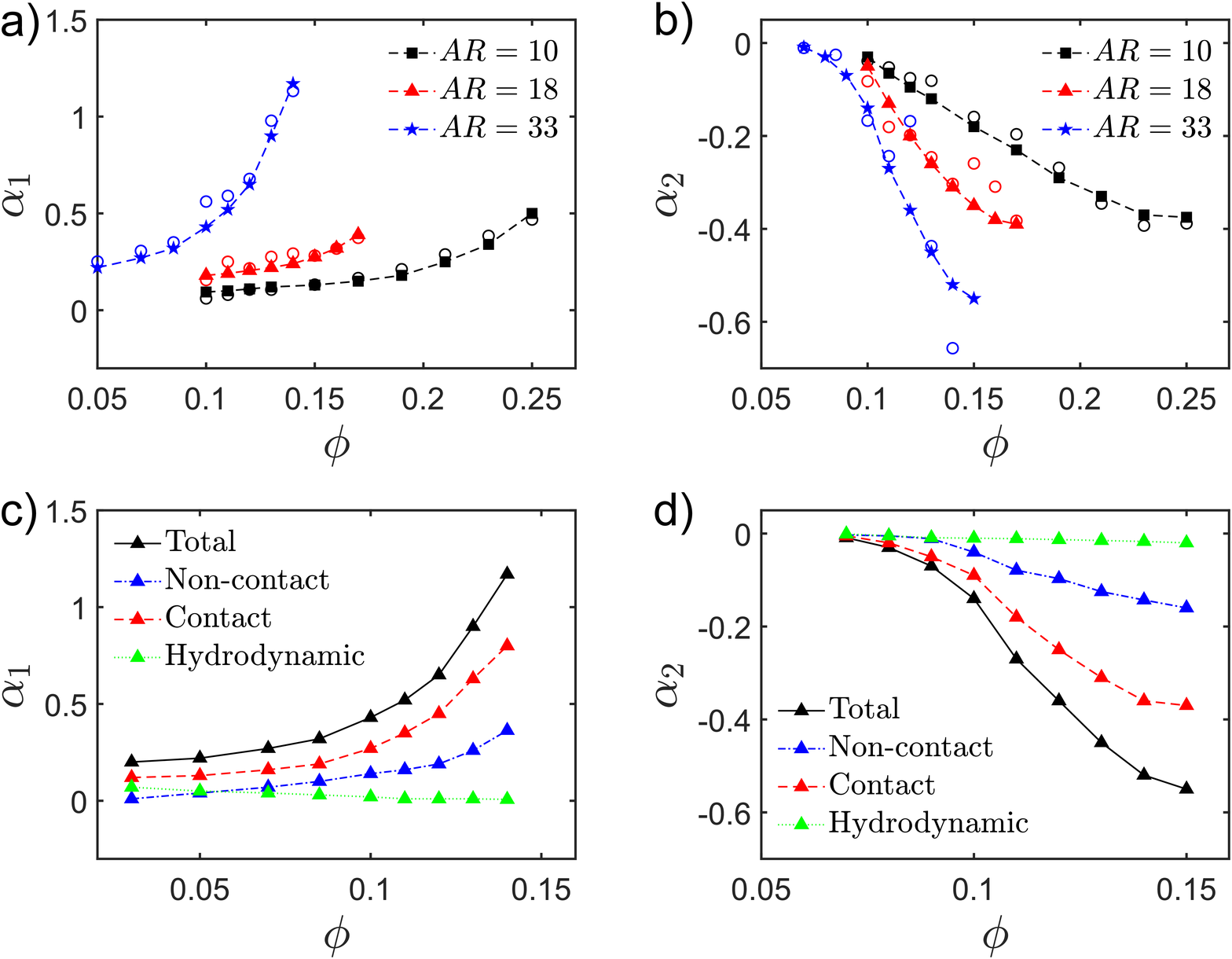}}
  \caption{Numerical and experimental comparison of (a) the first normal stress  coefficient $\alpha_1$ and (b) the second normal stress  coefficient $\alpha_2$   as a function of volume fractions for $AR$ = 10, 18 and 33. Hollow circles correspond to experimental data. Contributions arising from hydrodynamic,
  conservative, and contact forces to total (c) $\alpha_1$, (d) $\alpha_2$ are plotted as a function volume fractions. $\alpha_1$ is positive and $\alpha_2$ is negative in agreement with the experiments \citep{keshtkar2009rheological,snook2014normal,bounoua2016normal}. 
  The shear rate was fixed to $\dot{\gamma}/\dot{\gamma_0}= 30$ when the suspension viscosity reached a plateau. The data plotted here demonstrates that the contact
  contribution is the dominant contributor to the normal stress coefficients at higher shear rates.} 
  \label{fig:normal_stress_contribution}
\end{figure}

In fiber suspensions, normal stress differences inevitably arise due to the presence of hydrodynamics and inter-fiber interactions  \citep{keshtkar2009rheological}. 
We compute the first and second normal stress coefficients for $AR$ = 10, 18, and 33, and compare them with experiments \citep{bounoua2016normal} as shown in figure~\ref{fig:normal_stress_contribution}. The materials and experimental conditions are kept the same as the ones used for the viscosity measurement \citep{bounoua2016shear}. 
Specifically, the experiments are carried out at a higher shear rate when the viscosity reached a plateau. That is why we fix the shear rates to $\dot{\gamma}/\dot{\gamma_0}= 30$ when the suspension viscosity reaches a plateau in the numerical simulations.  All the simulation parameters are the same as in the last section. 
While the first normal stress coefficient $\alpha_1$ is positive, the second normal stress coefficient $\alpha_2$ is negative; in quantitative agreement with the experiments \citep{bounoua2016apparent, snook2014normal,keshtkar2009rheological}.  Moreover, the magnitude of $\alpha_2$ is smaller than that of $\alpha_1$. 

In addition, both normal stress coefficients decrease in magnitude with decreasing the aspect ratio, in agreement with experimental studies \cite{snook2014normal, bounoua2016apparent}. The origin of the normal stress coefficients at a high shear rate can be better understood by separating the hydrodynamic, contact, and non-contact (attractive+repulsive) contributions to the normal stress coefficient as shown in figures~\ref{fig:normal_stress_contribution}c and  \ref{fig:normal_stress_contribution}d. The hydrodynamic contribution to the normal stress coefficient is negligible for the conditions under consideration.  The data presented here clearly show that normal stress coefficients are primarily affected by contact. The normal stress coefficients of suspension increase in magnitude as the volume fraction rise due to an increase in both the contact and non-contact contributions. While the non-contact contribution increases weakly, the contact contribution becomes more significant at a higher volume fraction.  
Moreover, to investigate the shear rate dependencies, we measure the first and second normal stress coefficients for $AR$ = 18 and 33, as a function of f shear rate. We observe a similar shear rate dependence for the normal stresses as we do for the viscosity, and the data can be found in  supplemental material figure~S5. 

\subsection{Effect of coefficient of friction on the normal stress coefficients}
\begin{figure}
  \centerline{\includegraphics[width=1.0\linewidth]{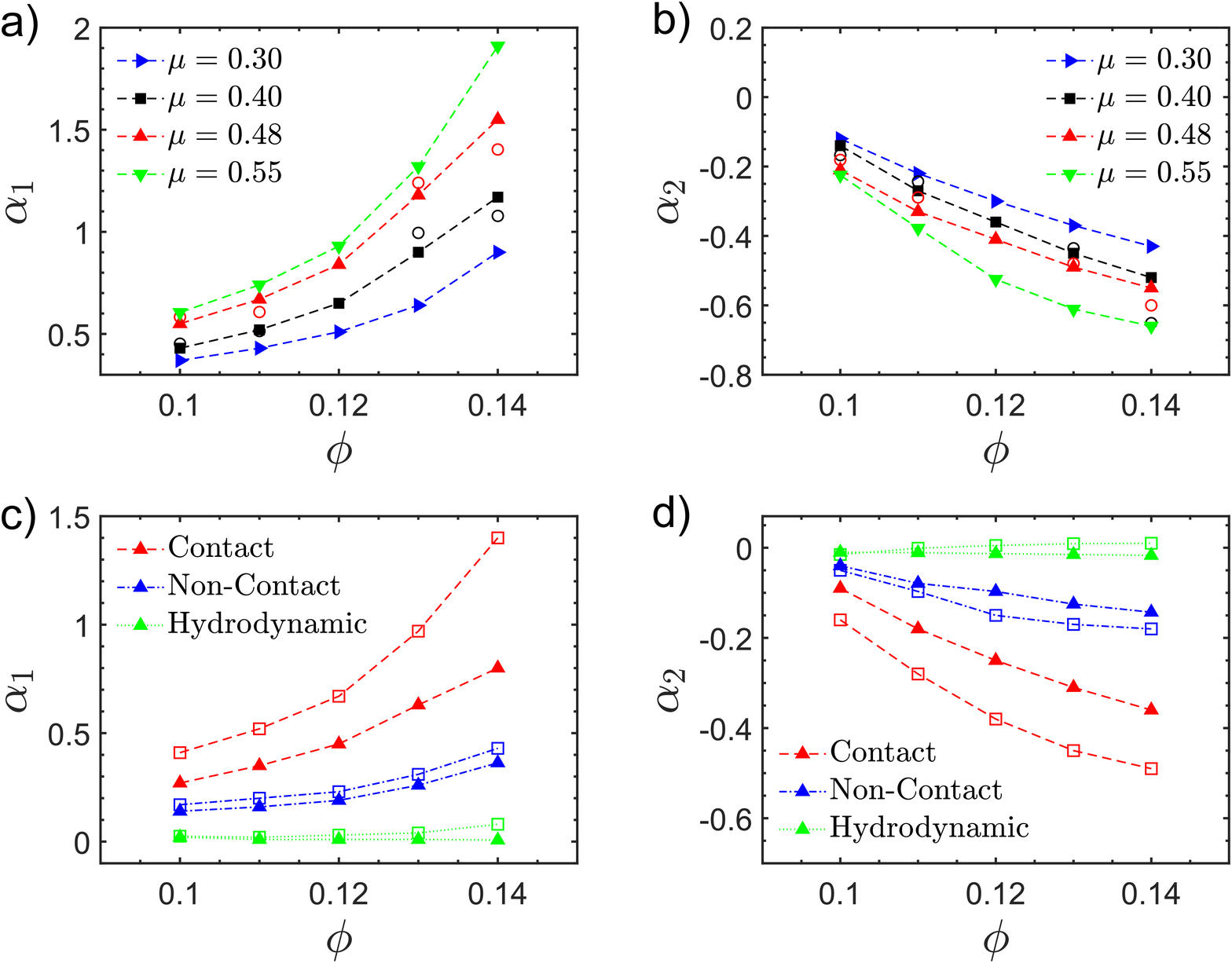}}
 \caption{(a) First normal stress coefficients $\alpha_1$, and (b) second normal stress coefficient  $\alpha_2$ as a function of volume fraction $\phi$ for different values  of friction coefficient $\mu$. Filled symbols with dashed lines show the numerical data. The black and red hollow circles show experimental values for unwashed and washed fibers, respectively. The aspect ratio of the fibers was fixed to $AR$ = 33. The effect of friction on the normal stress coefficients is significant, with a rise in $|\alpha_1|$ and $|\alpha_2|$. Hydrodynamic, contact and non-contact contributions to (c) first normal stress coefficient $\alpha_1$ and (d) second normal stress coefficient  $\alpha_2$ as a function of volume fraction $\phi$ for friction coefficient $\mu = 0.40$ (filled symbols) and  $\mu = 0.55$ (hollow symbols). As the volume fraction increases, the contact contribution increases significantly, leading to an increase in the magnitude of the normal stress coefficients.}
\label{fig:N1_N2_friction}
\end{figure}
We have provided a  quantitative explanation of the yield stress, normal stresses, and shear thinning in the fiber suspensions. Our model does an excellent job of reproducing the experimental data. This model can also capture and explain the effect of changing different parameters that control rheology. As an example, we will show the capability of the model to quantify the effect of changing fiber surface properties. 



We employed our model to quantify the effect of changing fiber surface properties and compare the results against the experimental observations by Bounoua \textit{et al.} \citep{bounoua2016apparent}. They changed the surface properties by washing the fibers. We hypothesize  that changing the fiber surface properties will directly modify the solid contact that can be  numerically quantified by varying the coefficient of friction. We measure the normal stress coefficients for five different volume functions ($\phi = 0.10, 0.11, 0.13, 0.14, 0.15$) with aspect ratio $AR$ = 33, varying the coefficient of friction $\mu$ between 0.30 to 0.55. Numerical results, having $\mu$ = 0.40 and 0.48, match closely with the experiment for washed and unwashed fibers, respectively, as shown in figures~\ref{fig:N1_N2_friction}a and \ref{fig:N1_N2_friction}b . These results demonstrate the applicability of the proposed model to capture and quantify the effect of modifying fiber surface properties.

The effect of friction is  more  pronounced  on $|\alpha_1|$ compared to  $|\alpha_2|$. To understand the underlying mechanism of the effect of modifying solid contact between fibers, we examine the hydrodynamic, contact, and non-contact contributions to the normal stress coefficients for $\mu$ = 0.40 (filled symbols) and $\mu$ = 0.55 (hollow symbols) in figures~\ref{fig:N1_N2_friction}c and \ref{fig:N1_N2_friction}d. We observe that the hydrodynamics contribution is almost independent of the friction coefficient. While the non-contact contribution increases weekly, the contact contribution increases significantly for the first normal stress coefficient $\alpha_1$. The noted increase in the magnitude of $\alpha_1$ is, therefore, solely due to the increase in the contact contribution. For the second normal stress coefficient $\alpha_2$ in figure~\ref{fig:N1_N2_friction}d, a similar conclusion holds: the friction weakly affects the hydrodynamic and non-contact contributions while it significantly increases the contact contribution.

\section{Conclusion}
Our research provides a fundamental insight into the complex rheological behavior of fiber suspensions based on balances between hydrodynamic, conservative, and contact forces. In this work, we provided the first quantitative explanation of the origin of yield stress, shear thinning, and normal stress differences in fiber suspensions by focusing on contact and non-contact contributions. Comparing the relative viscosity, the first normal stress coefficient $\alpha_1$, and the second normal stress coefficient $\alpha_2$ with experimental measurements corroborate the proposed model.  We demonstrated that the attractive interaction of van der Waals origin results in yield stress. Moreover, we explored the divergence of the yield stress as the suspension volume fraction, $\phi$
approaches the maximum flowable limit, $\phi_y^m$. Re-scaling volume fraction, $\phi$ by the jamming volume fraction, $\phi_y^m$ collapsed the yield stress data for varying aspect ratios on a single curve, denoting that changing the fiber aspect ratio affects the maximum volume fraction at which the yield stress diverges. In addition, we demonstrated that the degree of shear thinning and yield stress depends on the strength of attractive force, which can be regulated in principle by fiber size, microstructure, chemistry at solid-fluid interfaces, and fluid and solid phase parameters such as dielectric properties \citep{scales1998shear, galvez2017dramatic}. 

The first and second normal stress coefficients were compared with experiments for different aspect ratios and volume fractions to explore the dilute, semi-dilute, and concentrated regimes. The results showed that the first normal stress coefficient $\alpha_1$ is positive and the second normal stress coefficient $\alpha_2$ is negative, in agreement with available experiments \citep{keshtkar2009rheological,bounoua2016normal, snook2014normal}. As expected, the contact contribution is the dominant contribution to the normal stress differences at higher shear rates because non-contact contributions become less and less important as the shear rate increases.

A direct comparison of normal stress coefficients with the experiment was performed when the fiber surface was modified. We captured the impact of changing fiber properties by modifying the frictional contact by changing the coefficient of friction.  Interestingly, the friction appeared to act primarily through the contact stresses, as the hydrodynamic and non-contact stresses were unaffected by friction. 
Our results demonstrated the importance of accurately modeling the inter-fiber interactions to capture the experimentally observed shear rate-dependent rheological
behavior of fiber suspensions. 
Due to the complexity of attractive and friction forces at the microscopic scale and the number of parameters that are potentially relevant in the surface physical chemistry, the next step would be to quantitatively determine the attractive and frictional force from colloidal probe AFM measurements. 

\textbf{ACKNOWLEDGEMENT}

AMA would like to acknowledge financial support from the Department of Energy via grant EE0008910. 

\bibliography{paper}

\end{document}


\preprint{APS/123-QED}

\title{Supplementary Information to: Rheology of dense  fiber suspensions: Origin of yield stress, shear thinning and normal stress differences}



\author{Monsurul Khan}
 \affiliation{%
 Department of Mechanical Engineering, Purdue University, IN 47905, USA
}
\author{Rishabh V. More}%
\affiliation{%
 Department of Mechanical Engineering, Purdue University, IN 47905, USA
}%
\author{Luca Brandt}

\affiliation{
Linn´e Flow Centre and SeRC (Swedish e-Science Research Centre), KTH Mechanics, SE 100 44 Stockholm, Sweden
}
\author{Arezoo M. Ardekani}
\affiliation{Department of Mechanical Engineering, Purdue University, IN 47905, USA
}%

\date{\today}

\begin{abstract}
In this document we provide details about the numerical method used in this simulations. We provide further evidence in support of our model by comparing high shear rate viscosity for a range of volume fractions. At the end we report the shear rate dependent normal stresses in fiber suspension. 
\end{abstract}
\maketitle

\vspace{0.5cm}
\centerline{\large{\textbf{Numerical method}}}

\begin{figure}
  \centerline{\includegraphics[width=0.7\linewidth]{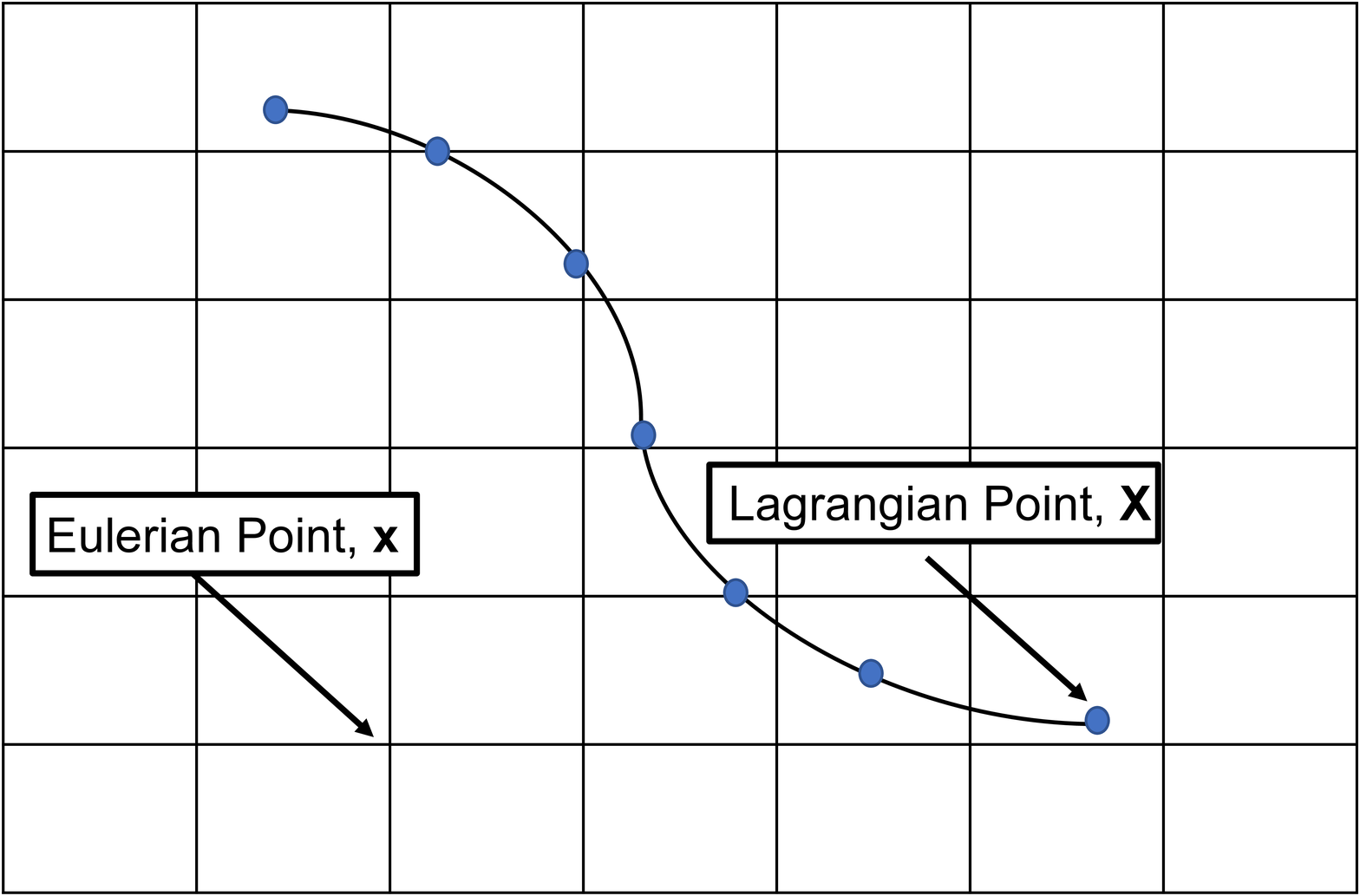}}
  \caption{Schematic of the Eulerian and the  Lagrangian grids. The blue dots denote the Lagrangian points through which the position of the fibers are  defined.}
\label{fig:eulerian}
\end{figure}

The fluid-solid coupling is achieved using the Immersed Boundary Method (IBM) \citep{peskin1972flow}. In the IBM, the geometry of the object is represented by a volume force distribution $\mathbf{f}$ that mimics the effect of the object on the fluid. In this method, two sets of grid points are needed: a fixed Eulerian grid $\mathbf{x}$ for the fluid and a moving Lagrangian grid $\mathbf{X}$ for the flowing deformable structure as shown in figure \ref{fig:eulerian}. Each fiber has its own Lagrangian coordinate system. 
In this study we assume the fibers to be neutrally buoyant, so equation of motion of fiber is 
\begin{equation}
\frac{{{\partial ^2}\mathbf X}}{{\partial {t^2}}} = \frac{{{\partial ^2}{\mathbf X_{fluid}}}}{{\partial {t^2}}} + \frac{\partial }{{\partial s}}(T\frac{{\partial \mathbf X}}{{\partial s}}) - B\frac{{{\partial ^4}\mathbf X}}{{\partial {s^4}}} - \mathbf F + \mathbf {F}^f,
\label{pinal}
\end{equation}
where the LHS term is the acceleration of the fiber, and the RHS consists of the acceleration of the fluid particle at the fiber location and the different forces acting on the fibers. Here, $s$ is the curvi-linear coordinate along the fiber, $\mathbf X = (x(s,t),y(s,t),z(s,t))$ is the position of the Lagrangian points on the fiber axis, $T$ is the tension, $B$ is the bending rigidity,  $\mathbf F$ is the fluid-solid interaction force, $\mathbf {F}^f$ is the net inter-fiber interaction. To solve for the fiber position in equation \ref{pinal}, we first solve the Poisson's equation for tension  as:
\begin{equation}
\frac{{\partial \mathbf X}}{{\partial s}}.\frac{{{\partial ^2}}}{{\partial {s^2}}}(T\frac{{\partial \mathbf X}}{{\partial s}}) = \frac{1}{2}\frac{{{\partial ^2}}}{{\partial {t^2}}}(\frac{{\partial \mathbf X}}{{\partial s}}.\frac{{\partial \mathbf X}}{{\partial s}}) - \frac{{{\partial ^2}\mathbf X}}{{\partial t\partial s}}.\frac{{{\partial ^2}\mathbf X}}{{\partial t\partial s}} - \frac{{\partial \mathbf X}}{{\partial s}}.\frac{\partial }{{\partial s}}(\mathbf{F}^a + \mathbf{F}^b + \mathbf{F}^f - \mathbf F),
\label{Poission}
\end{equation}
where $\mathbf{F}^a = \frac{{{\partial ^2}\mathbf{X_{fluid}}}}{{\partial {t^2}}}$ is the acceleration of the fluid particle at the fiber location and $\mathbf{F}^b =  - B\frac{{{\partial ^4}\mathbf{X_{}}}}{{\partial {s^4}}}$ is the bending force. The effect of the moment exerted by the fluid on the freely suspended fibers also appears in equation \ref{Poission} and has been discussed elsewhere \citep{banaei2020numerical}.


As the fibers are freely suspended in the fluid medium, we impose zero force, torque and tension at the free ends.
\begin{equation}
    \frac{{{\partial ^2}\mathbf X}}{{\partial {s^2}}} = 0, \frac{{{\partial ^3}\mathbf X}}{{\partial {s^3}}} = 0, \textrm{ and } T = 0.
\label{BC}    
\end{equation}
At each time step, the fluid velocity is first interpolated onto the Lagrangian grid points using the smooth Dirac delta function, $\delta$ \citep{roma1999adaptive}:
\begin{equation}
   {\mathbf U_{ib}} =  \int\limits_V {{\mathbf  u(\mathbf x,t)\delta (\mathbf X - \mathbf x)dV}}, 
\end{equation}
The fluid and solid equations are then coupled by the fluid-solid interaction force, 
\begin{equation}\label{eq:FSI_force}
    \mathbf F = \frac{{\mathbf U - {\mathbf U_{ib}}}}{{\Delta t}},
\end{equation}
where  $\mathbf U_{ib}$ is the interpolated fluid velocity on the Lagrangian points defining the fibers, $\mathbf U$ is the velocity of the Lagrangian points and $\Delta t$ is the time step. 
The Lagrangian force is then extrapolated onto the fluid grid by
\begin{equation}
    \mathbf f(\mathbf x,t) = \frac{\pi}{4}r_p^2\int\limits_L {\mathbf  F(\mathbf x,t)\delta (\mathbf X - \mathbf x)ds}.
\end{equation}
Here $r_p = d/l$, is the slenderness ratio of the fiber, which is the inverse of its aspect ration defined as $AR=l/d$. 

\vspace{0.5cm}
\centerline{\large{\textbf{Lubrication interactions}}}

To accurately resolve the lubrication interactions between the fibers when the inter-fiber gap falls below a few grid-sizes, we use the lubrication correction model of \cite{lindstrom2008simulation}. The lubrication force model is based on two infinite cylinders for two different cases: the cylinders can be parallel or at an arbitrary angle. The first-order approximation of the lubrication force for the non-parallel case was derived by \cite{yamane1994numerical} and is given as follow:
\begin{equation}
    \mathbf F_1^l = \frac{{ - 12}}{{{\mathop{\rm Re}\nolimits} \sin \alpha }}\frac{{\dot{\mathbf{h}}}}{h},
\label{lub1}    
\end{equation}
here $h$ is the shortest distance between the cylinders, $\dot{\mathbf{h}}$ is the relative normal velocity between the closest points on the fibers, and $\alpha$ is the contact angle. The first order approximation of the lubrication force per unit length between parallel cylinders was derived by \cite{kromkamp2005shear}:
\begin{equation}
    \begin{array}{l}
\mathbf F_2^l = \frac{{ - 4}}{{\pi {\mathop{\rm Re}\nolimits} r_p^2}}\left( {{A_1} + {A_2}\frac{h}{a}} \right){\left( {\frac{h}{a}} \right)^{ - 3/2}}\dot{\mathbf{h}},\\
{A_1} = 3\pi \sqrt 2/8, {A_2} = 207\pi \sqrt 2 /160,
\end{array}
\label{lub2}
\end{equation}
here $a$ is the cylinder radius $(a = d/2)$. Based on equations \ref{lub1} and \ref{lub2}, the following approximation of the lubrication force for two finite cylinders is \citep{lindstrom2008simulation}:
\begin{equation}
    {\mathbf F^l} = \min \left( {\mathbf F_1^l/\Delta s,\mathbf F_2^l} \right).
\end{equation}
The force for the non-parallel case is divided by the Lagrangian grid spacing $\Delta s$ to calculate the force per unit length.
The numerical method implemented to calculate the lubrication force is discussed in \cite{banaei2020numerical}, and hence is not repeated here. As the distance between the fibers becomes the order of the mesh size, hydrodynamic interactions are not well resolved; to address this issue, we introduce a lubrication correction force \citep{lindstrom2008simulation}. When the shortest distance between two Lagrangian points becomes lower than $d/4$, we introduce lubrication correction force as: $\mathbf F^{lc}=\mathbf F^{l}-\mathbf F_0^l$, where $F_0^l$ is the lubrication force at a $d/4$ distance. 

The lubrication forces diverge as the minimum inter-fiber separation decreases, and theoretically should prevent the fibers from coming into direct contacts. However, the thin lubrication film between close fibers can break because of the presence of irregularities on their surfaces leading to a direct contact between the fibers and hence, contact forces. 

\vspace{0.5cm}

\centerline{\large{\textbf{Stress and bulk rheology calculations}}}

The bulk stress in the suspension is required to quantify the rheological properties of the suspension. The total stress in the suspension in terms of contributions from hydrodynamics and fiber stresses in the dimensionless form is \citep{wu2010method}:
\begin{equation}
    \Sigma {_{ij}}  = {\mathop{\rm \textit{Re}}\nolimits} \left[ {\frac{1}{V}\int\limits_{V - \sum {{V_f}} } {\left( { - p{\delta _{ij}} + \frac{2}{{Re}}{e_{ij}}} \right)dV + \frac{1}{V}\sum\limits_1^n {\int\limits_{{V_f}} {{\sigma _{ij}}} dV} }  - \frac{1}{V}\int\limits_V {{{u'}_i}} {{u'}_j}dV} \right].
\end{equation}
Here, $V$ is the total volume, $V_f$ is the volume occupied by each fiber, ${e_{ij}} = \frac{{\partial {u_i}}}{{\partial {x_j}}} + \frac{{\partial {u_j}}}{{\partial {x_i}}}$ represents the strain rate tensor, and $\boldsymbol{u}'$ is the velocity fluctuation. $\sigma_{ij}$ is the fiber stress \citep{batchelor1971stress}. The dimensionless total stress consists of the fluid bulk stress and the stress generated by the presence of fibers and the interactions between them. So, the total stress (${\Sigma}_{ij}$) can be written as
\begin{equation}
     {\Sigma}_{ij} = {\Sigma}{_{ij}^0}+{\Sigma}{_{ij}^f},
\end{equation}
\begin{equation}
     \begin{array}{l}
{\Sigma}{_{ij}^0}  = \frac{{{\mathop{\rm \textit{Re}}\nolimits} }}{V}\int\limits_{V - \sum {{V_f}} } {\left( { - p{\delta _{ij}} + \frac{2}{{{\mathop{\rm \textit{Re}}\nolimits} }}{e_{ij}}} \right)} dV,\\
\Sigma _{ij}^f = {\textit{Re} \over V}\sum \int\limits_{{V_f}} {{\sigma _{ij}}} dV - {{{\mathop{\rm \textit{Re}}\nolimits} } \over V}\int\limits_V {{{u'}_i}} {{u'}_j}dV.
\end{array}
\end{equation}
Here, ${\Sigma}_{ij}^0$ is the viscous fluid stress and results in a dimensionless contribution of 1 (or $\eta\dot{\gamma}$ in the dimensional form) in a simple shear flow after subtracting the isotropic fluid pressure. ${\Sigma}{_{ij}^f}$ is the stress generated by the presence of fibers and inter-fiber interactions. The fiber stress $\sigma_{ij}$ can be decomposed into two parts:
\begin{equation}
\int\limits_{{V_f}} {{\sigma _{ij}}} dV = \int\limits_{{A_f}} {{\sigma _{ik}}} {x_j}{n_k}dA - \int\limits_{{V_f}} {\frac{{\partial {\sigma _{ik}}}}{{\partial {x_k}}}{x_j}} dV,
\end{equation}
where $A_f$ represents the surface area of each fiber and $\mathbf{n}$ is the unit surface normal vector on the fiber pointing outwards. The first term is called the stresslet, and the  second  term  indicates  the  acceleration  stress \citep{guazzelli2011physical}. The second term is identically zero for neutrally buoyant fibers when the relative acceleration of the fiber and fluid is zero. $\sigma _{ik}{n_k}$ is simply the force per unit area acting on the fibers \citep{batchelor1971stress}. Hence, for slender bodies, $\sigma _{ik}{n_k}$ can be rewritten as:

\begin{equation}
\int\limits_{{A_f}} {{\sigma _{ik}}} {x_j}{n_k}dA =  - r_p^2\int\limits_L { {F}_i} {x_j}ds,
\end{equation}
${F}_i$ is the fluid solid interaction force as defined in equation~\ref{eq:FSI_force}. The term $r_p^2$  arises from choosing the linear density instead of the volume density in the characteristic force scale. Finally, the total fiber stress is defined as:
\begin{equation}
    \Sigma {_{ij}^f}  = - \frac{{Rer_p^2}}{V}\sum\limits_1^n\int\limits_L {{F_i}{x_j}ds}  - \frac{{{\rm{\textit{Re}}}}}{V}\int\limits_V {{{u'}_i}} {{u'}_j}dV.
\end{equation}
From the results of our simulations, we observe that, the last term related to the velocity fluctuations are very small compared to the stresslet and can be neglected for the range of Reynolds number considered here. The calculated bulk stress tensor can now be used to quantify rheological properties of the suspensions. 

There are three main contributions to the bulk stress: 1) the hydrodynamic contribution $\Sigma_{ij}^h$, 2) the contact contribution $\Sigma_{ij}^c$, and 3) the non-contact contribution $\Sigma_{ij}^{nc}$.  The contact and non-contact contribution can be calculated from the ensemble average of the contact and non-contact stresslet respectively given by:

\begin{equation}
\begin{array}{l}
   \Sigma_{ij}^c =  -\frac{Rer_p^2}{V}\sum\limits_1^n \int_{L}{{F}^c_i} {x_j}ds,\\
   \Sigma_{ij}^{nc} =  -\frac{Rer_p^2}{V}\sum\limits_1^n \int_{L}{{F}^{nc}_i} {x_j}ds.
   \end{array}
\end{equation}
Thus, the hydrodynamic contribution can be simply obtained as $\Sigma_{ij}^h=\Sigma_{ij}-\Sigma_{ij}^c-\Sigma_{ij}^{nc}$. This splitting of the total stress allows us to track the variations in the contributions from different mechanisms to the observed rheological behavior of the suspension with varying parameters, e.g., $\eta_r^h=\Sigma_{xy}^h$, $\eta_r^c=\Sigma_{xy}^c$, and  $\eta_r^{nc}=\Sigma_{xy}^{nc}$ are the hydrodynamic, the contact and non-contact contributions to the relative viscosity $\eta_r$, respectively.

\vspace{0.5cm}

\centerline{\large{\textbf{Relative viscosity at high shear rate}}}

\begin{figure}
  \centerline{\includegraphics[width=1.0\linewidth]{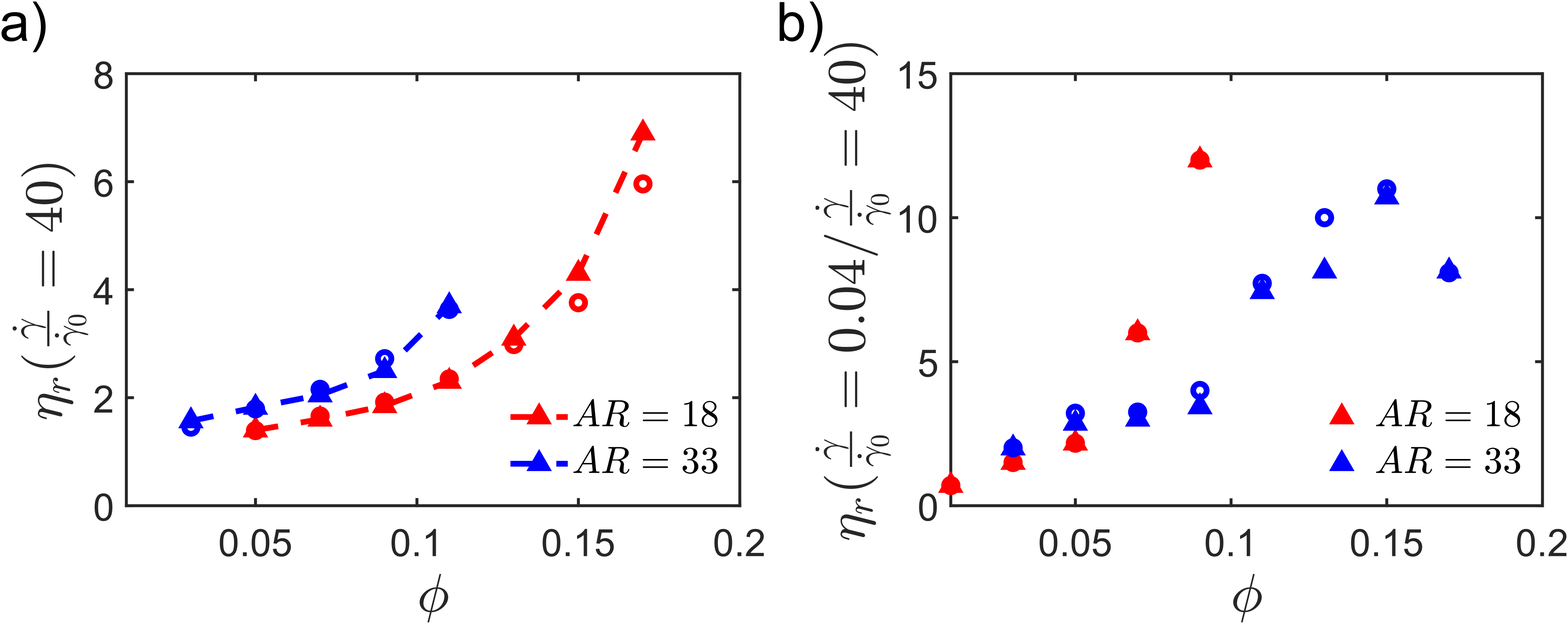}}
  \caption{Experimental and numerical comparison of the (a) high shear viscosity, and (b) ratio of low-shear to high-shear viscosity as  a function of suspension volume fractions. Hollow circles corresponds to experimental data.}
\label{fig:viscosity_ratio}
\end{figure}

In order to better understand the concentration effect on the low shear viscosity, the ratio of low shear viscosity to the high shear viscosity is plotted as a function of fibers volume fraction in figure~\ref{fig:viscosity_ratio}b for both aspect ratios. For both suspensions, the viscosity ratio increase as fiber volume fraction increases, indicating that the suspensions are significantly more shear thinning at high concentrations than at low concentrations. 

\begin{figure}
  \centerline{\includegraphics[width=0.5\linewidth]{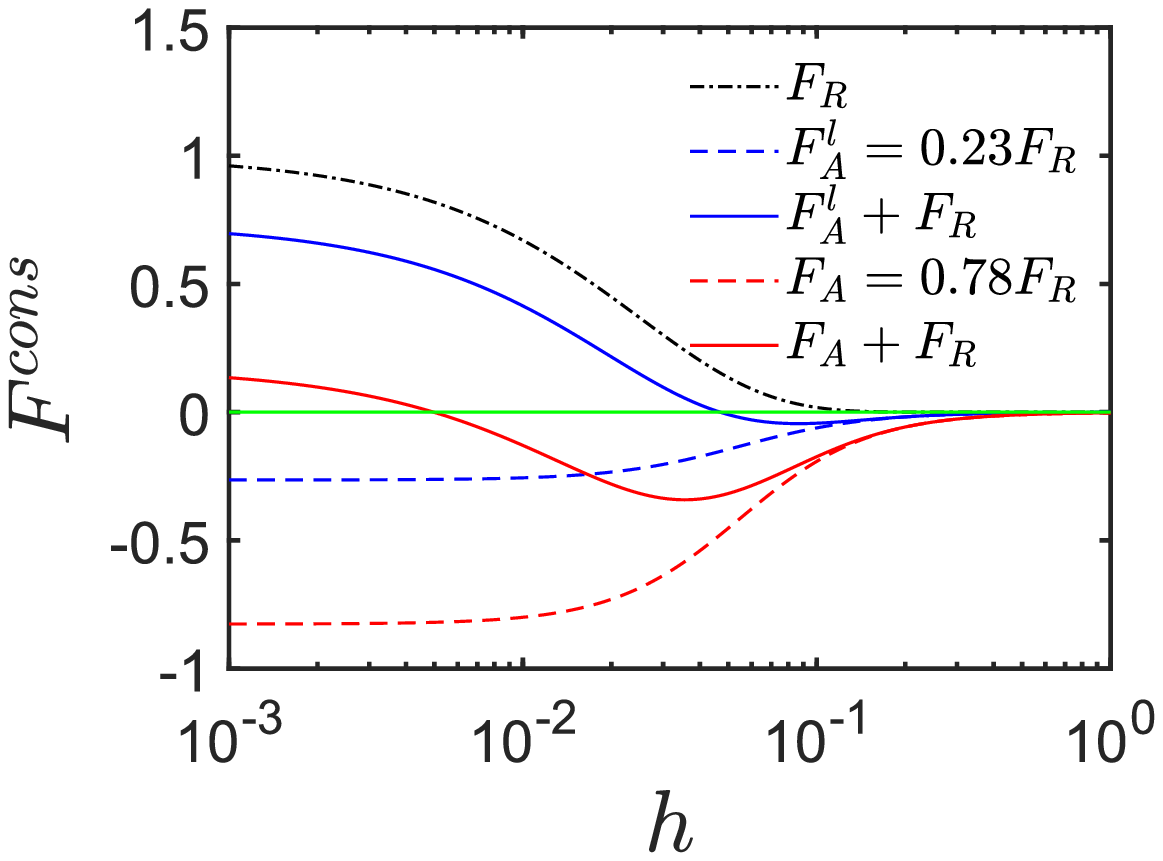}}
  \caption{Conservative force profiles as a function of surface separation for two different attractive
force magnitudes. The horizontal green line shows the zero value.}
\label{fig:conservative}
\end{figure}

We plot the conservative force profiles as a function of separation distance for two attractive forces $F_A$ in figure~\ref{fig:conservative}. With the increase in $F_A$, the separation distance $h$ below which the conservative force is repulsive decreases. At a sufficiently high attractive force, the position of zero force and cut-off separation for lubrication ($d/4$) are the same, bringing fibers into direct frictional contact. These frictional contacts can withstand applied shear stress, increasing the yield stress and viscosity.



\vspace{0.5cm}

\centerline{\large{\textbf{Shear rate dependencies of normal stress differences}}}

\begin{figure}
 \centerline{\includegraphics[width=1.0\linewidth]{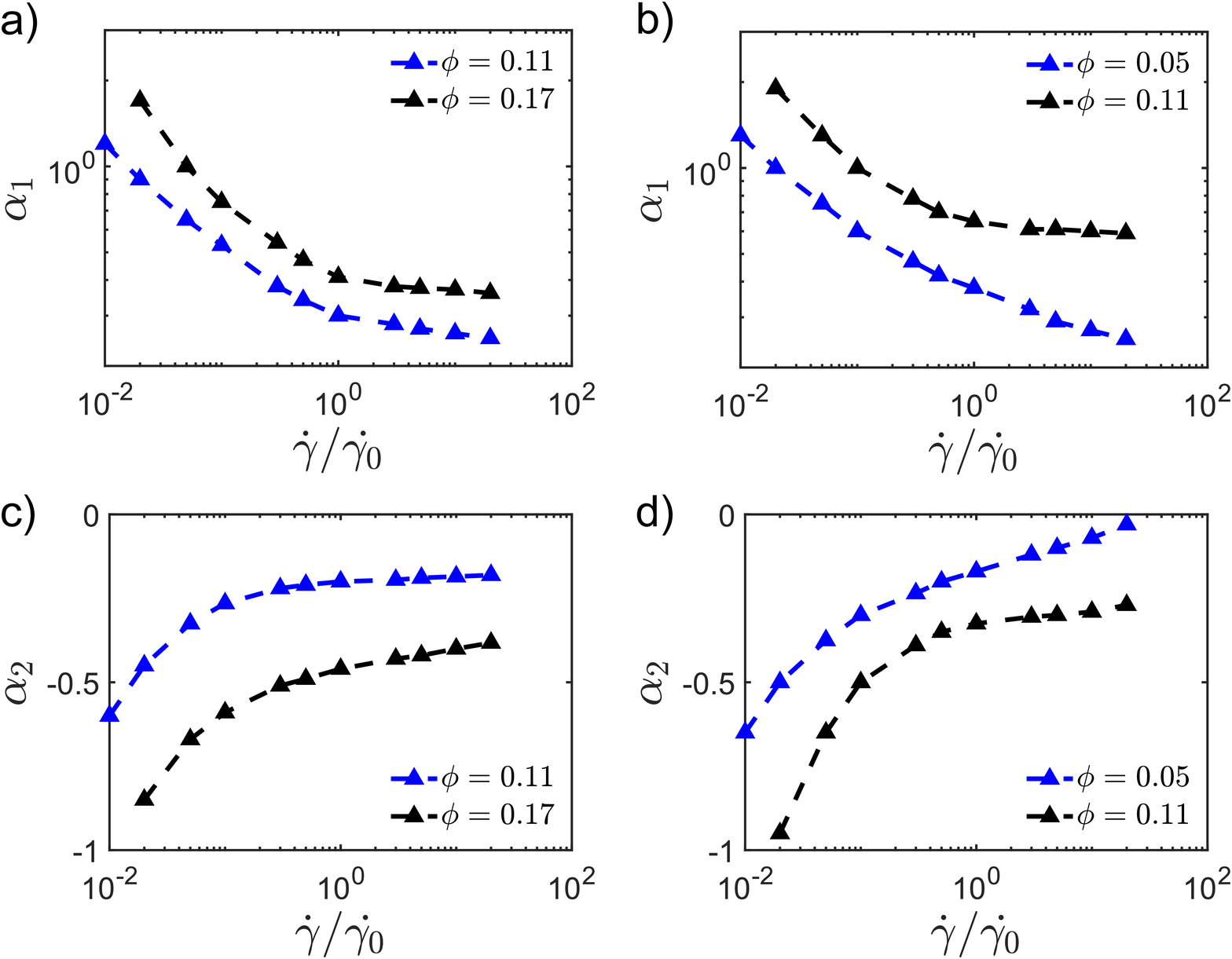}}
  \caption{Evolution of first normal stress coefficient $\alpha_1$ with applied dimensionless shear rate for aspect ratios (a) $AR = 18$, (b) $AR = 33$; $\alpha_1$ is always positive.  Evolution of second normal stress coefficient $\alpha_2$ with applied dimensionless shear rate for aspect ratios (c) $AR = 18$, (d) $AR = 33$; $\alpha_2$ is always negative.}
\label{fig:normal_shear}
\end{figure}

To investigate the shear rate dependencies, the first and second normal stress coefficients are plotted in figure~\ref{fig:normal_shear} for two different aspect ratios, $AR$ = 18 and 33 as a function of dimensionless shear rate.  The magnitude of normal stress coefficients shows a shear thinning behavior similar to viscosity. 

\vspace{0.5cm}

\bibliography{supplemental}
